\documentclass{elsart}

\usepackage{epsfig,bm,epsf,float,amssymb}

\begin{document}

\begin{frontmatter}



\title{Experimental Results from an Antineutrino Detector for Cooperative Monitoring of Nuclear Reactors}

\author[sandia]{N.~S.~Bowden\corauthref{cor1}},
\corauth[cor1]{Corresponding Author. Tel.: +1 925 294 2566.}
\ead{nbowden@sandia.gov}
\author[llnl]{A.~Bernstein},
\author[sandia]{M.~Allen},
\author[sandia]{J.~S.~Brennan},
\author[llnl]{M.~Cunningham},
\author[sandia]{J.~K.~Estrada},
\author[sandia]{C.~M.~R.~Greaves\thanksref{thanks1}},
\author[llnl]{C.~Hagmann},
\author[sandia]{J.~Lund},
\author[sandia]{W.~Mengesha},
\author[sandia]{T.~D.~Weinbeck\thanksref{thanks2}},
\author[llnl]{C.~D.~Winant}

\address[sandia]{Sandia National Laboratories, Livermore, CA~94550, USA}
\address[llnl]{Lawrence Livermore National Laboratory, Livermore, CA~94550, USA}

\thanks[thanks1]{Present Address: Lawrence Berkeley National Laboratory, Berkeley, CA~94720, USA}
\thanks[thanks2]{Present Address: Department of Physics and Astronomy, Tufts University, Medford, MA~02155, USA}

\begin{abstract}
Our collaboration  has designed, installed, and operated a compact antineutrino
detector at a nuclear power station, for the purpose of monitoring the power
and plutonium content of the reactor core. This paper focuses on the basic
properties and performance of the detector. We describe the site, the reactor
source, and the detector, and provide data that clearly show the expected
antineutrino signal. Our data and experience demonstrate that it is possible to
operate a simple, relatively small, antineutrino detector near a reactor, in a
non-intrusive and unattended mode for months to years at a time, from outside
the reactor containment, with no disruption of day-to-day operations at the
reactor site. This unique real-time cooperative monitoring capability may be of
interest for the International Atomic Energy Agency (IAEA) reactor safeguards
program and similar regimes.

\end{abstract}

\end{frontmatter}

\section{Introduction}
\label{sec:intro}

The IAEA uses an ensemble of procedures and technologies, collectively referred
to as the Safeguards Regime, to detect diversion of fissile materials from
civil nuclear fuel cycle facilities into weapons programs. The research
described here concerns safeguards for one important element of this cycle,
nuclear reactors. Earlier work~\cite{rovno,firstpaper} has described a method
for exploiting the large number ($\approx {10^{21}~\bar{\nu}}$/s) of
antineutrinos emitted by fission reactors to track the plutonium content and
thermal power of the reactor core in real time. This approach offers unique
advantages that are complimentary to existing safeguards methods. The potential
impact on the regime is large, inasmuch as over $200$ of the world's civil
power reactors, located in some $30$ countries, are subject to IAEA
safeguards~\cite{iaearef}.

To be useful for reactor safeguards, antineutrino detectors must be simpler to
construct and operate than the current generation of detectors, which were
built to investigate the basic physics of the neutrino sector. It is in this
respect that this work differs from the experimental effort described
in~\cite{rovno}. Furthermore, an experimental deployment at a commercial
reactor is essential to demonstrate the potential of this technology to the
global non-proliferation community.

In this paper, we describe the characteristics and performance of such a
device; a simple, compact prototype antineutrino detector, ``SONGS1'', which
has been installed at a commercial nuclear power station to demonstrate the
feasibility of this monitoring technique.

\section{Reactor Monitoring with Antineutrino Detectors}

In the context of nuclear nonproliferation, cooperative monitoring refers to
the tracking of nuclear inventories and nuclear facilities with the agreement
of the host country. The IAEA Safeguards Regime is the most prominent example
of such monitoring. Antineutrinos have unique features that make them
interesting for the purpose of cooperative monitoring of nuclear reactors. The
highly penetrating nature of the particle and the high flux near reactors means
that relatively small detectors external to the reactor can acquire a signal
useful for monitoring, even with hour to day integration times. Furthermore,
over a broad range of antineutrino energies, the ratio of antineutrinos per
fission and MeV emitted by uranium and plutonium varies by factors of
$50$--$90\%$~\cite{vogel2}. As a result, as uranium is consumed and plutonium
produced in the core, the measured antineutrino rate and spectrum will change,
providing up to date information about the burnup and isotopic content of the
core.

For example, during a typical $1$--$1.5$~year cycle at a pressurized water
reactor (PWR), the net effect on the antineutrino rate due to the fissile
isotopic evolution is $6$--$12\%$. Information about the reactor's fissile
content is a key element of the IAEA safeguards regime, whose purpose is to
detect unauthorized diversion of fissile material from civil nuclear fuel
cycles. Antineutrino detectors can supply this information in real time and in
a manner wholly under the control of the safeguards agent, without interfering
with daily reactor operations. Because of these interesting features,
antineutrino detection holds promise as a complementary technique for IAEA and
other reactor safeguards regimes.

\subsection{Relative and Absolute Measurements}
\label{sec:relative}

In neutrino physics experiments that require an absolute flux or spectral
measurement, considerable effort must be devoted to a full understanding of the
absolute efficiency of the detector. In safeguards regimes, both relative and
absolute measurements may be of use. Relative measurements, in which the
antineutrino rate or spectrum is compared to some initial value, are
insensitive to time-independent systematic shifts in the detector response,
such as an error in an estimate of the detection efficiency. For such
measurements, a detailed understanding of the individual detector's efficiency
is therefore not required, so long as the detector response is stable in time.
In the analysis presented here, we have estimated the efficiency of the SONGS1
detector using a simple detector simulation. We emphasize that the potential
utility of the detector for safeguards does not depend on a precise  knowledge
of this efficiency.


\section{Site and Reactor Characteristics}
\label{sec:site}

The SONGS1 detector is deployed at Unit 2 of the San Onofre Nuclear Generating
Station (SONGS). There are two operational reactors at this station; both are
pressurized water reactors designed by Combustion Engineering in the 1970s and
have maximum thermal (electric) power of $3.4$~GW$_{th}$ ($1.1$~GW$_{elec}$).

\begin{figure}[tb]
\centering
\includegraphics*[width=3in]{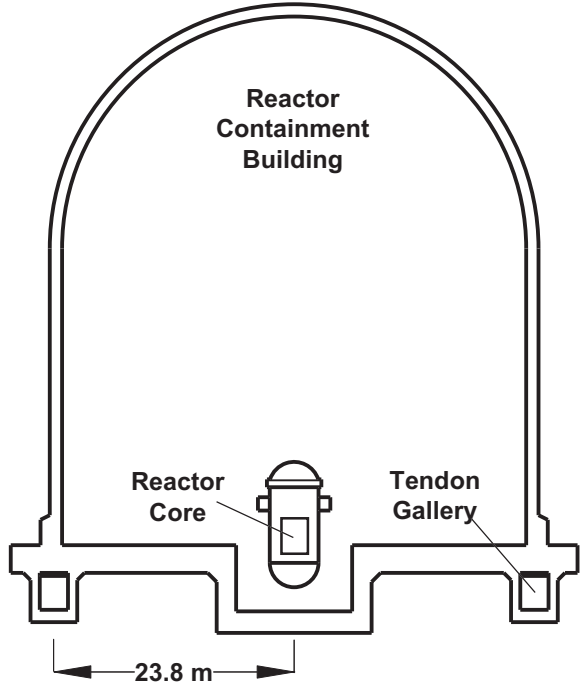}
\caption{A cross section of the SONGS Unit 2 containment building showing the
position of the reactor vessel and tendon gallery.} \label{fig:SONGS}
\end{figure}

The detector is located in the tendon gallery of Unit~2 (Fig.~\ref{fig:SONGS}).
A feature of many commercial
reactors~\cite{tendon_paper1,tendon_paper2,tendon_paper3,tendon_paper4,tendon_paper5},
a tendon gallery is an annular concrete hall that lies beneath the walls of the
reactor containment structure. It is used to inspect and adjust the tension in
reinforcing steel cables which extend throughout the concrete of the
containment structure. At SONGS these inspections occur every two to three
years, and involve the examination of only a handful of representative tendons.
Subsequently, the placement of the detector in the tendon gallery has little or
no impact on regular plant operations.

SONGS1 is located $24.5\pm1.0$~m from the Unit~2 reactor core and $149\pm3$~m
from that of Unit 3. Since the antineutrino flux generated by each core is
isotropic, $97$\% of the reactor antineutrinos reaching the SONGS1 detector
originate from Unit~2. With Unit~2 at full power the antineutrino flux at the
SONGS1 location is $\approx 10^{17}~$m$^{-2}$s$^{-1}$.

It is worth noting that at this position the flux of other particles, in
particular fast neutrons and gamma rays, reaching the detector from the reactor
core is practically zero. For example, the material between the core and the
detector corresponds to nearly $100$~attenuation lengths for $10~$MeV neutrons,
compared with a total flux exiting the reactor that is no greater than
$10^8~$cm$^{2}$s$^{-1}$. This yields a vanishingly small reactor neutron flux
in the gallery. Similar arguments apply for reactor produced gamma rays.

The tendon gallery is located about $10$~m below the surface, providing
$\approx25$~meters of water equivalent (mwe) overburden which eliminates the
soft component of the cosmic ray flux and reduces the muonic component by a
factor of $7$~\cite{Miyake}. An important purpose of this prototype deployment
was to determine whether cosmic ray induced backgrounds at this relatively
shallow depth would overwhelm the reactor antineutrino signal in a simple
detector.

The temperature in the tendon gallery is not actively controlled. Due to the
underground location diurnal temperature variations at the detector are small
($< 0.1~^\circ$C), although there is a slow seasonal variation of up to
$5~^\circ$C. Small drifts in detector gain that result from this variation are
accounted for via the detector calibration procedure
(Sec.~\ref{sec:calibration}).

In the context of cooperative monitoring, tendon galleries are ideal locations
for antineutrino detectors. They are remote from daily reactor operations,
provide significant overburden, and are as close as possible to the core while
remaining outside of containment.

\section{Antineutrino Interactions in Liquid Scintillator}
\label{sec:nubarint}

Antineutrinos are detected via the inverse beta decay process on quasi-free
protons in hydrogenous scintillator: $\bar{\nu}_e + p \rightarrow e^{+} + n$.
The positron ($e^{+}$) and neutron ($n$) are detected in close time
coincidence, allowing strong rejection of the much more frequent singles
backgrounds due to natural radioactivity or neutron production. The positron
deposits energy via Bethe-Bloch ionization as it slows in the scintillator, and
upon annihilation with an electron, emits two gamma rays which can deposit up
to an additional $1.022$~MeV of energy. Because the ionization and annihilation
gamma ray energy depositions are effectively simultaneous, and promptly follow
the antineutrino interaction, this combination of energy depositions is
measured as a single event in the detector, referred to as the ``prompt''
energy deposition. Assuming complete containment of the event, the prompt
energy and the antineutrino energy are directly related through the equation:

\begin{equation}
E_{\nu} = E_{prompt} + 1.8~MeV - 2\cdot m_{e}.
\label{eq:nubar_energy}
\end{equation}

In the current detector, this ideal relation is distorted by escape of
$0.511~$MeV annihilation gamma rays that are created near the edge of the
detector. Monte Carlo transport simulations account for this loss of energy,
and allow for a full analysis of the deposited energy spectrum as well as
providing an estimate of the efficiency for detection of the prompt energy
(Sec.~\ref{sec:positron}).

The neutron carries away a few keV of energy from the antineutrino
interaction~\cite{vogel_direction} and is detected by capture on a Gadolinium
(Gd) dopant.  The liquid scintillator is doped at $0.1\%$ by weight with
natural Gd, which is chosen for its very high neutron capture cross-section.
The Gd doping yields a neutron capture time of $28~\mu$s and an energy release
of $\approx 8~$MeV via a gamma ray cascade, compared to a capture time of
$200~\mu$s and energy release of $2.2~$MeV for capture on hydrogen. The
interactions in the detector resulting from this $8~$MeV cascade are referred
to as the ``delayed'' energy deposition. As with the positron, the measured
energy in the detector is somewhat less than $8~$MeV since some of the cascade
gamma rays escape the detector volume without depositing their full energy.
Still, the increase in deposited energy and the decrease in the time interval
combine to significantly improve the overall background rejection capability
relative to capture in undoped scintillator.

We can therefore expect to observe two classes of events in a detector based
upon Gd loaded liquid scintillator. The first, referred to here as correlated
events, involve a prompt energy deposition associated with a neutron that is
captured a short time later on a Gd nucleus. The time separation between prompt
and delayed energy depositions will follow an exponential distribution with
time constant equal to the neutron capture time of the Gd doped scintillator
($28~\mu$s). The second class, referred to as uncorrelated backgrounds,
involves the random ``coincidence'' of two independent energy depositions
caused by natural backgrounds. We refer to these individual energy depositions
as ``singles.'' Since these backgrounds obey Poisson statistics, the time
separation between singles will also follow an exponential distribution with
time constant equal to the inverse of the singles rate (effectively the
detector trigger rate).

Unfortunately, antineutrino interactions are not the only events that produce
correlated events in the detector. For example, fast neutrons, produced by muon
capture or muon spallation \cite{fast_neutron,fast_neutron2,fast_neutron3}, can
scatter off protons in the scintillator, giving a prompt energy deposition, and
then be captured on Gd with the characteristic time distribution. Therefore,
observation of correlated events alone is not sufficient to measure the reactor
antineutrino flux -- the correlated background rate must also be determined.
Accurate \textit{a priori} calculations of the background are difficult since
they depend strongly upon the geometry and composition of the detector
surroundings. In practice, one must wait for a reactor off period to measure
the correlated background rate.


\section{SONGS1 Detector Description}
\label{sec:detector}

The SONGS1 detector consists of three subsystems; a central detector containing
the Gd-doped liquid scintillator target and photomultiplier tubes, a passive
water or polyethylene shield on all sides, and a plastic scintillator muon veto
system placed outside the water shield on five sides of the detector.

In addition to the detector itself the installation at SONGS includes the Data
Acquisition System~(DAQ, Sec.~\ref{sec:daq}) and a computer that automatically
performs the data analysis described in Sec.~\ref{sec:analysis}. A modem is
used to automatically retrieve the results of this analysis as well as detector
state of health indicators, allowing for the remote monitoring of the detector,
and indeed of reactor operation. The SONGS1 system typically operates without
physical intervention for months at a time.

\subsection{Central detector}
\label{sec:cells}

The central detector, seen in Fig.~\ref{fig:detector}, consists of four
identical stainless steel cells, each with inner dimensions of $\approx 43~$cm
by $\approx 43~$cm by $\approx 98~$cm, filled with $\approx 0.64~$tons of
liquid scintillator. Stainless steel was used rather than the more typical
acrylic to avoid any possibility of attack by the scintillator on the container
walls, and subsequent leakage of the flammable scintillator. To improve
reflectivity, the sides and bottom of each cell are lined with foil wrapped
acrylic sheets. To increase the fraction of totally internally reflected (TIR)
light, these sheets were sealed within transparent FluorinatedEthylenePropylene
(FEP) bags filled with Argon~\cite{Bugey}. The bags make it difficult to
measure the detector volume to high precision, resulting in a $\approx 10\%$
uncertainty in this quantity. However, this is of no consequence for our
relative measurement of the reactor antineutrino flux.

We use scintillator remaining after the completion of the Palo Verde
antineutrino detection experiment~\cite{paloverde}. This scintillator has a
density of $0.856~$g~cm$^{-3}$ and is $13\%$ hydrogen by weight. The
development and characteristics of this scintillator are described in detail
elsewhere~\cite{paloverde_scint}.

Scintillation light generated by the interaction of particles in the cells
propagates to $9~$inch hemispherical Photonis~XP1802 photomultiplier tubes
(PMTs) by means of an acrylic coupling cylinder submerged in the liquid and
extending through an opaque acrylic lid. There are two PMTs mounted on each
cell (not shown in Fig.~\ref{fig:detector}).

\subsection{Passive Shielding}
\label{sec:pass_shielding}

Both gamma ray and neutron fluxes in the central detector are reduced by
passive water or polyethylene shielding which surrounds the detector on six
sides. The water is contained in acrylic tanks, supported by perforated angle
stainless steel racks. On average the passive layer is $\approx 0.5$~m thick.
An MCNP simulation predicts that a shield of this thickness reduces the flux of
$10~$MeV neutrons and $2~$MeV gamma rays by factors of $25$ and $10$
respectively.

\subsection{Muon Veto}
\label{sec:active_shielding}

\begin{figure}[tb]
\centering
\includegraphics*[width=3in]{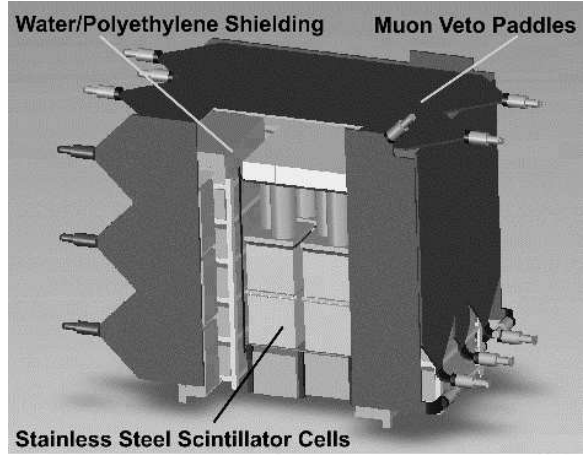}
\caption{A cut away diagram of the SONGS1 detector, showing the major
subsystems.} \label{fig:detector}
\end{figure}

A $2~$cm thick plastic scintillator envelope identifies and rejects cosmic ray
muons. As shown in Fig.~\ref{fig:detector}, the envelope is comprised of
scintillator paddles of different sizes placed on five sides of the detector,
read out by photomultiplier tubes. The paddles are wrapped first with aluminum
foil to enhance reflectivity and then with black tape for light tightness.

For the top, horizontally oriented paddles, the well separated muon peak was
identified in the paddle energy distribution, at its expected location of
$5~$MeV. The energy threshold defining a muon was set three sigma below this
peak, with the sigma defined  by a fit to the rising lobe of the
Landau-distributed muon spectrum. This procedure introduces a small amount of
contamination from the tail of the paddle gamma ray distribution, which extends
above $3~$MeV, but the total deadtime introduced by the muon veto even with
this contamination remains small. Due to the geometry of muon interactions, the
muon spectrum in the vertically oriented paddles has no clear peak. Thresholds
in vertically oriented paddles were set relative to the rates in the horizontal
paddles, by assuming a $cos^2(\theta)$ distribution for the muon flux, which
predicts that the flux through a vertically oriented paddle should be $50\%$
that of a horizontally oriented paddle. The total rate at which muons are
recorded by this system is $\approx 500$~Hz.

\section{Expected Antineutrino Rates}
\label{sec:expected_rate}

The number of detected antineutrino interactions in the detector is directly
related to the individual isotopic fission rates through the equation

\begin{equation}
N_{\bar{\nu}}  = \left(\frac{T N_{p}}{4 \pi D^{2}}\right) \sum_{i} {f_{i}} \int
dE_{\bar{\nu}} \sigma(E_{\bar{\nu}}) \phi_{i}(E_{\bar{\nu}})
\epsilon(E_{\bar{\nu}}). \label{eq:nurate}
\end{equation}

In this equation, $N_{\bar{\nu}}$ is the number of detected antineutrino
interactions in a time interval T. $N_{p}$ is the number of target protons in
the detector, $D$ is the distance from the detector to the center of the
reactor core, $f_{i}$ is the number of fissions per second from the $i$th
isotope. In practice the index runs over the four dominant fissioning isotopes
$^{235}$U, $^{238}$U, $^{239}$Pu and $^{241}$Pu, which account for $\approx
99.9\%$ of fissions in a typical reactor. $\sigma$ and $\phi_{i}$, both
depending on the antineutrino energy $E_{\bar{\nu}}$, are respectively the
cross-section for the inverse beta interaction, and the number density per MeV
and fission for the $i$th isotope. $\epsilon$ is the energy dependent intrinsic
detection efficiency. The integral extends over all antineutrino energies above
the $1.8~$MeV threshold of the inverse beta interaction.

With our current $0.64\pm0.06~$ton liquid scintillator detector at a standoff
of $24.5\pm1$~m from the reactor core, and with the reactor parameters defined
in Sec.~\ref{sec:site}, the rate predicted by Eq.~\ref{eq:nurate} at the
beginning of the reactor fuel cycle is approximately $3800\pm440$ antineutrino
interactions per day for a perfectly efficient detector.


\section{Electronics and Data Acquisition System}
\label{sec:daq}

\begin{figure*}[tb]
\centering
\includegraphics*[width=6.5in]{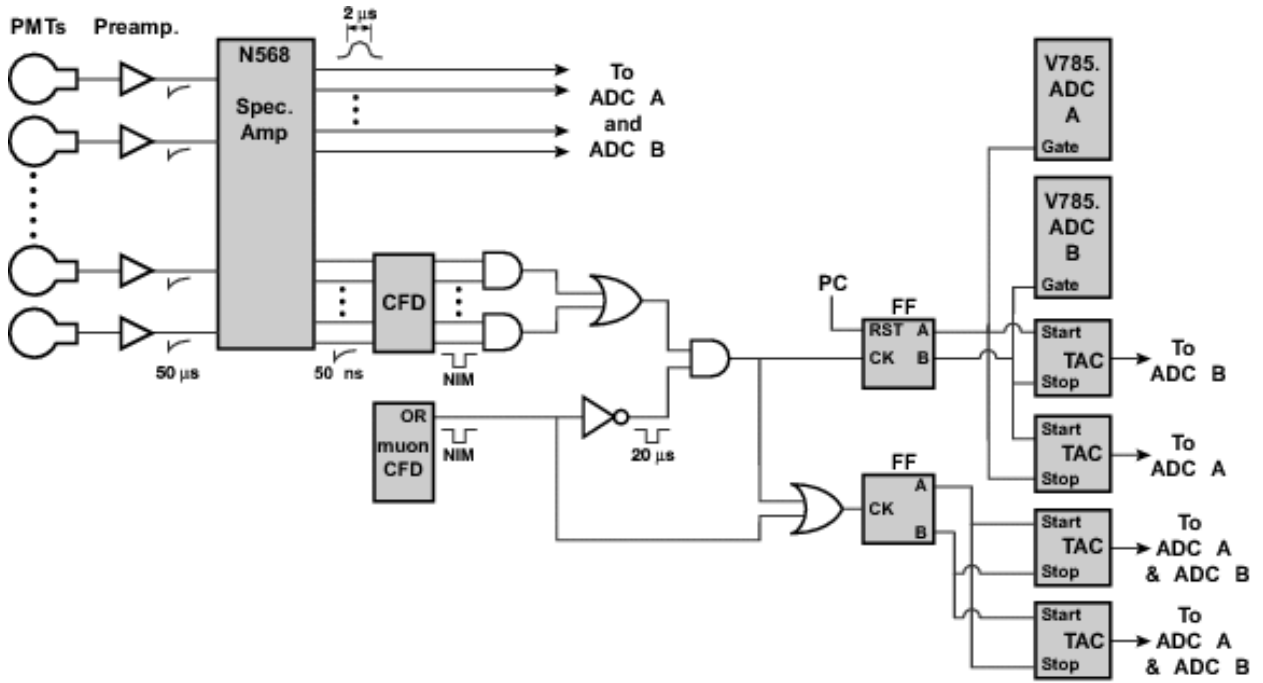}
\caption{ The data acquisition system and trigger logic.} \label{fig:schematic}
\end{figure*}

Fig.~\ref{fig:schematic} shows the architecture of the data acquisition and
trigger. As mentioned earlier, there are two PMTs per scintillator cell. Custom
preamplifiers built into the bases of the PMTs produce tail pulses that are
sent to a CAEN~N568 spectroscopy amplifier for shaping. The approximately
Gaussian shaped output of the N568 is passed directly to two CAEN~V785 peak
sensing Analog-to-Digital-Converters (ADCs).

A fast amplified signal from the N568~amplifier is passed to a CAEN~V812
Constant Fraction Discriminator (CFD) module. Hardware thresholds are set at
$\approx 1.5$~MeV for each PMT, so as to allow online calibration using the
$2.6~$MeV $^{208}$Tl gamma ray from the Thorium
chain~(Sec~\ref{sec:calibration}).

The anode outputs of all muon paddles are passed directly to CAEN V812 CFDs,
and the OR output from these is used for vetoing and timing purposes.

Gates for the ADCs are produced by the following trigger logic. A coincidence
is formed from the two discriminator outputs corresponding to the PMTs on the
same cell using an ORTEC~C4020 logic module (a cell trigger -- typically
$120~$Hz per cell). The logical OR of the four cell triggers is taken
(typically $500~$Hz), and combined with the logical NOT of a $20~\mu$s long
muon veto signal. A non-retriggerable logic pulse of $3~\mu$s duration (an
acquisition trigger -- typical rate $420~$Hz) is produced at this stage since
this is the approximate duration of the shaped output of the N568; accepting
events with shorter separation would only result in pulse pileup at the ADCs.

Thus, an event that causes two PMTs from any single cell to fire in
coincidence, at least $20~\mu$s after the passage of the last muon and $3~\mu$s
after the last cell trigger, will result in one of the ADCs being gated. Gates
are generated for each ADC in turn, so that while one ADC is digitizing (V785
digitization time $\approx6~\mu$s) the other can accept another event. Gates
are generated until the $32$~event buffer on each ADC is full. Once these
buffers have been emptied, the flip flop generating the alternating gates is
reset so that it is known which ADC receives the first trigger, allowing
subsequent events to be formed into correctly ordered pairs in later
analysis~(Sec.~\ref{sec:analysis}).

In similar fashion, pairs of Time-to-Amplitude-Converters (TACs) are used to
record timing information. The ORTEC~566 TACs utilized have a long $50~\mu$s
reset time, which if used singly would result in an unacceptably large fraction
of events being recorded without timing information. Thus, start and stop
signals are alternated between a pair of TACs; the stop for one is the start
for the other. To record the time between pairs of events (the interevent time)
the acquisition trigger serves as both the start and the stop signal. To be
certain that the combination of ADC and TAC deadtimes does not introduce a
non-uniform response, we only consider interevent times greater than $10~\mu$s
in the analysis that follows.

To record the time to last muon, the acquisition trigger serves as the stop
signal, while the start signal is either the previous acquisition trigger or a
muon signal if that occurred sooner. This start logic is necessary since the
TACs are single shot -- the production of a TAC output pulse clears the timer.
Always recording the time to last event, either muon or cell, solves this
problem, but introduces about $4$\% extra deadtime. Both TAC signals are
digitized simultaneously with the PMT pulses by the peak sensing ADCs.


\section{Detector Simulation}

\label{sec:MC}

A high fidelity detector simulation is not necessary for the relative
antineutrino rate measurements of interest here. Indeed, this lack of reliance
on simulation is a strength in the context of the practical application of
antineutrino detection technology to safeguards. Instead, we used Monte Carlo
simulations to gain insight into the atypical particle transport properties of
the SONGS1 detector, and to estimate the efficiency of our detector.

We used the MCNPX~\cite{MCNP} package, which simulates particle transport, but
does not model the transport of the optical photons generated in the
scintillation process. To account for the optical photon response, we used a
simple model that incorporates photostatistics and the measured vertical
position variation in each cell. Because the variation in light collection has
an important effect on the response function, a  fuller understanding of the
detector response ultimately requires a simulation or measurement of the
transport properties of visible scintillation photons throughout the entire
detector volume. However, even without these inputs, we can still estimate the
energy scale and the positron and neutron detection efficiencies, and confirm
the expected transport behavior of positrons, neutrons and gamma rays in the
detector and surrounding materials.

\subsection{The Detector Simulation in MCNP}

We simulated particle transport in the scintillator, the surrounding materials
including the acrylic optical couplings, the PMTs, the reflective panels inside
the detector, the stainless steel detector walls and support structure, the
water shield and the plastic muon veto. Characteristic gamma rays for the
background simulation (Sec.~\ref{sec:gamma_bkg}) were propagated from a maximum
depth of $0.5~$m in the surrounding concrete -- beyond this depth there was a
negligible chance for interaction in the detector.

\subsection{The Vertical LED Scan and the Detector Optical Model}
\label{sec:opticalmodel}

\begin{figure}[tb]
\centering
\includegraphics*[width=3in]{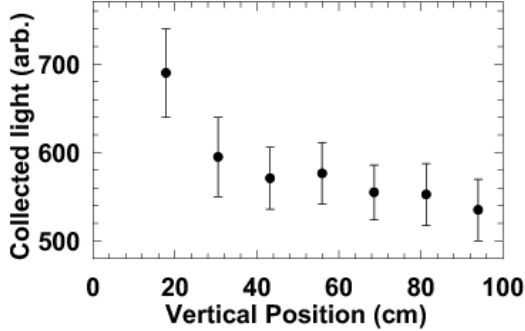}
\caption{The average PMT response as a function of vertical position in a
single cell. The position is measured with respect to the top of the cell.}
\label{fig:LEDfig}
\end{figure}

Partial characterization of the optical response of the detector was derived
from a vertical scan in one cell with a light emitting diode (LED). The LED
position was offset approximately 8 centimeters transversely from the center of
the cell. Pulse height spectra were recorded at seven positions from
$10$~to~$90$~cm from the top of the scintillator cell. The variation in
response, averaged over the two PMTs, is plotted in Fig.~\ref{fig:LEDfig}. As
seen in the figure, the response is fairly uniform until the source is within
about $20~$cm of the top of the cell.

We use this data set to parameterize the effects of both photostatistics and
vertical position on the detector energy resolution. The parametrization is
then used to estimate the effect of optics on the energy resolution in the
gamma ray, neutron, and positron transport simulations described below. We
model the optical resolution function $\sigma_{opt}(E)$ as:

\begin{equation}
\sigma_{opt}(E) = \sigma_{pos}(E)+ \sigma_{p.e.}(E) = k1\cdot E + k2 \cdot
\sqrt{E} \label{eq:opticalres}
\end{equation}

Here $\sigma_{pos}(E)$ represents the average effect of the vertical position
on the PMT response, and $\sigma_{p.e.}(E)$ is the variation in response caused
by the finite number of photoelectrons generated in the event. A value of
$0.03$ for $k1$ is estimated from the LED data in Fig.~\ref{fig:LEDfig}. $k2$,
accounting for photostatistics, has a value of $0.15$, which is the typical
fractional width of the LED pulse hight spectra in the position scan.

Although horizontal variation of light collection has an important impact on
the resolution, a horizontal scan at different heights was impractical and was
not included in the model. While this ultimately limits the fidelity of the
simulation, this simple model is sufficient to estimate the energy scale and
detection efficiencies, and confirms our main expectations about the detector
response to antineutrino events.

\subsection{Gamma-ray Energy Calibration}
\label{sec:gamma_bkg}

As described in Sec.~\ref{sec:calibration} the detector energy calibration is
derived from a fit to a spectral feature due to the $2.6~$MeV $^{208}$Tl gamma
ray from the Thorium chain and to measured ADC pedestal values. This broad
feature lies between the Tl photopeak and Compton edge, and arises from the
multiple Compton interactions that a gamma ray is likely to undergo in the
detector. The size of the detector is such that there is a high likelihood that
a large fraction of the gamma ray energy will be lost within it. To estimate
the average energy deposited by this process, and thus the correct energy to
assign the measured position of this feature in our calibration process, we
propagated gamma rays from natural Uranium (U), Thorium (Th) and Potassium (K)
in external concrete through the detector components just described and into
the active detector volume.

Gamma interactions from U, Th, and K are all generated independently, and a
weighted sum of the resulting spectra is used to form the simulated total
gamma-ray energy spectrum. In comparing data and Monte Carlo, the relative
isotopic abundances of U, Th, and K are taken as free parameters, along with an
overall scale factor. Smearing due to position and photostatistics is fixed by
the optical model defined in Sec.~\ref{sec:opticalmodel}. The four free
parameters are optimized by fitting to the data in the $1$~to~$3~$MeV gamma ray
calibration energy range. The $\chi^2$ per degree of freedom for the fit is
$1.2$.

\begin{figure}[tb]
\centering
\includegraphics*[width=3in]{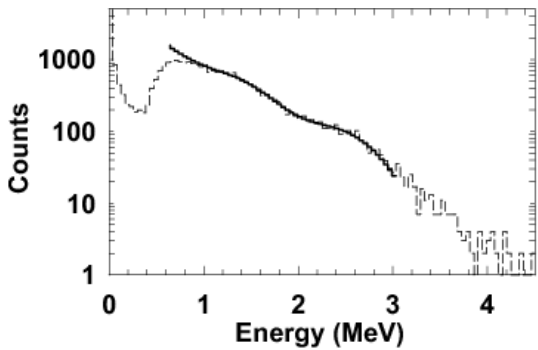}
\caption{External gamma interactions as recorded in the SONGS1 detector,
compared with the MCNP simulation (solid line).} \label{fig:mcdatagamcomp}
\end{figure}

Fig.~\ref{fig:mcdatagamcomp} compares the data and Monte Carlo. Once agreement
is optimized in the energy range of interest, the energy scale is set in the
data by matching the measured ADC values of the multiple Compton scatter
feature to the energies reported by the Monte Carlo. The inferred energy values
for the K and Th lines are $1.32$~and~$2.39~$MeV respectively.

Despite the absence of a full optical model, the simulation reproduces the main
features of the data, namely the spectral features due to the multiple Compton
scattering of K and Th gamma rays, and shows reasonable fidelity across the two
orders of magnitude variation in the pulse height spectrum.

\subsection{Positron Simulation}
\label{sec:positron}

Antineutrino-generated positrons, with energies from zero to several MeV,
travel a few centimeters from their point of creation and annihilate with the
creation of two $511~$keV gamma rays. We simulated positrons randomly
distributed throughout the liquid scintillator volume, and tallied the total
visible energy recorded in each cell, including the positron ionization energy,
and the energy deposited by scattering of the annihilation gamma rays.

The positron energy distribution is directly related to the antineutrino energy
distribution through Eq.~\ref{eq:nubar_energy} in Sec.~\ref{sec:nubarint}.
Because the positron energy spectrum changes as the reactor evolves, the
simulated spectrum must be integrated over the same burnup period as the actual
data. In the data presented here, this corresponds to the first $77$ days of
Cycle 14 of the SONGS reactor. The variation due to the changing isotopics in
the core is correctly accounted for in our predicted spectrum.

Fig.~\ref{fig:prompt_energy}(b) shows the simulated and measured positron
spectra in our detector. The fixed parameters from the optical model are used
to define the effects of position and photostatistics on the positron energy
scale. The only free parameter is an overall normalization factor.

The distributions reveal agreement at approximately the $5~\%$ level between
data and the model. A fuller treatment of the optical response is required for
a more precise comparison. We estimate a detection efficiency of $55\% \pm 5\%$
for antineutrino generated positrons in the prompt energy range of $2.39~$ to
$9~$MeV. The uncertainty is estimated by comparing the sum of the absolute
magnitudes of bin to bin count variations  between the data and Monte Carlo
spectra with the total number of data counts in the prompt energy range.

\subsection{Neutron Simulation}
\label{sec:neutron_mc}

Neutrons from inverse beta decay within the liquid scintillator are created
with energy of order $1$~keV. Subsequent elastic collisions thermalize the
neutrons until they are either captured or escape from the scintillator volume.
With a natural Gd loading of $0.1$\% by weight, approximately $88$\% of neutron
captures will be on Gd isotopes ($^{155}$Gd $82$\%, $^{157}$Gd $18$\%), with
the remainder capturing on protons.

The response of the detector to thermal neutrons generated by antineutrino
interactions has an important effect on the overall efficiency. We simulated
the gamma-ray cascade from neutron capture on Gd with the DICEBOX
algorithm~\cite{dicebox}. About $105$~gamma cascades per isotope with an
average multiplicity of $\approx5$ were calculated and written to a file. Our
modified version of MCNPX read in a new set of energies whenever a capture on
Gd occurred. The resulting gamma rays were then propagated isotropically from
the final neutron position, tracked, and energy deposits were tallied for each
detector cell. To estimate the efficiency for detection of neutrons from
antineutrino interactions, the simulated neutrons were distributed uniformly
throughout the active volume of the detector. Neutron energies were sampled
from the energy distribution of antineutrino-generated neutrons, and allowed to
thermalize and capture in the detector.

Fig.~\ref{fig:e_spectrum_extraction}(b) shows a comparison of the data and
Monte-Carlo, where the data are selected using all criteria that define an
antineutrino, as presented in Sec.~\ref{sec:analysis}. As with the positron
simulation, the only free parameter is an overall normalization constant.

Again, the agreement between data and Monte Carlo is reasonable considering the
simplicity of the model.  Qualitatively, the Monte Carlo confirms a key
expectation for this type of detector: there is a broad range of energy
depositions arising from the Gd capture, which extends well below the $8~$MeV
total energy of the gamma ray cascade. The peak near $2.2~$MeV due to capture
on hydrogen is seen in the simulation, as well as a large number of events that
deposit very little energy in the detector due to leakage.  The broad response
function to Gd captures is due to the fact that the shower spatial extent,
defined by the typical $1$--$2~$MeV energies of the gamma rays in the cascade,
is of the order of the detector size. This means that only those interactions
in the center of the detector will have a significant probability of depositing
all of their energy in the detector. Interactions away from the detector center
are likely to lose one or more gamma rays outside the detector volume,
resulting in a lower visible energy deposition and accounting for the broad
energy spectrum seen in the both the data and the simulation.

Based on the Monte Carlo, we estimate a detection efficiency of $40~\% \pm
4~\%$ for antineutrino generated neutrons in the delayed energy region of
$3.5~$ to $10~$MeV.

\subsection{Predicted Detection Efficiency}

\label{sec:totaleff}

Aside from the prompt and delayed energy thresholds, the detection efficiency
is also affected by the PMT asymmetry cut, the interevent and muon timing
selection cuts and the deadtime of the DAQ. As described in
Sec.~\ref{sec:analysis}, the interevent time distribution of antineutrino
interactions is described by an exponential decay. We estimate the efficiency
of this portion of our DAQ by integrating this exponential over the range of
accepted interevent times ($10$--$800~\mu$sec). The fractional dead time
imposed by the muon veto is given by the product of the $100~\mu$s width veto
with the sum of the trigger and muon rates (Sec.~\ref{sec:daq}). Our means of
estimating the fiducial volume of the detector, as defined by an analysis cut,
is described in Sec.~\ref{sec:event_selection}.

Table~\ref{tab:efftab} shows the estimated efficiencies through each of our
selection criteria, and the resulting total efficiency. It is interesting to
note that even with the$~40$--$50~\%$ efficiency through the positron and
neutron cuts, the underlying event rate is still high enough to give several
hundred antineutrino events per day in our detector. When combined with the
expected antineutrino interaction rate in our detector
(Sec.~\ref{sec:expected_rate}) this efficiency allows us to estimate the number
of antineutrino events that we should observe: $407\pm75$~/day.

\begin{table}
\caption{Detection efficiency.} \label{tab:efftab}
\begin{tabular}{l r r} \hline
Quantity       & & Efficiency    \\ \hline
Positron~~&~&~~$55\pm5$\%~    \\
Neutron~~&~&~~$40\pm4$\%~    \\
DAQ     &~&~~~$59\pm0.5$\%~  \\
\textit{-Livetime} &~~$92\pm0.5$\%~    &~  \\
\textit{-Muon veto} &~~$91\pm0.2$\%~    &~  \\
\textit{-$t_{min}=10~\mu$s } &~~$70\pm0.5$\%~    &~  \\
Fiducial      &~&~~~$83\pm4$\%~  \\ \hline Total &~&~~~$10.7\pm1.5$\%~
\\ \hline
\end{tabular}
\end{table}


\section{Detector Calibration}
\label{sec:calibration}

Energy and time calibrations are performed periodically for each PMT-ADC pair
and each TAC. The time scale is fixed by exercising each TAC with known
fixed-interval start-and-stop pulses created by a pulse generator.

\begin{figure}[tb]
\centering
\includegraphics*[width=3in]{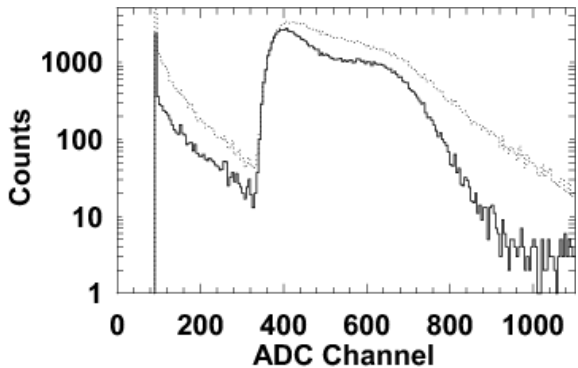}
\caption{A representative calibration spectrum from PMT 1, ADC 1 (solid). For
comparison, the spectrum that results from accepting PMT ratios as great as
$0.4$ is also shown (dotted).} \label{fig:th_spectrum}
\end{figure}

Relative calibration for each PMT-ADC pair and correction for long term drifts
depends primarily on the interaction of the $2.6~$MeV $^{208}$Tl gamma ray from
the Thorium chain with the detector. This gamma ray activity arises from the
concrete walls and floor of the tendon gallery. Multiple Compton scatters of
the gamma ray in the detector and surrounding material produce a broad spectral
feature in the measured pulse height spectrum (Fig.~\ref{fig:th_spectrum}).

The location of this feature, which is well fit by a gaussian, and the location
of the electronic pedestal value (defined as the ADC channel measured for zero
input voltage to the ADC) is automatically determined from each hour's data
file. Since the signals from all cells are digitized when any one cell
triggers, a large number of recorded events for a particular cell will have
zero energy, i.e. will be a measurement of the energy pedestal (negligible
electronic cross-talk is observed between PMTs). Calibration constants are
applied for each PMT-ADC pair separately. The calibration procedure is as
follows: determine the location of the energy pedestal; rebin the energy
spectrum from $4000~$channels to $800~$channels; apply the PMT ratio cut
described in Sec.~\ref{sec:analysis} with a ratio of $0.1$; fit the $^{208}$Tl
gamma ray feature to the sum of two exponentials and a Gaussian peak.

A more restrictive PMT ratio cut is used in the calibration procedure ($0.1$)
than in the analysis used to extract candidate antineutrino events ($0.4$).
While this means that the calibration samples only a subset ($\approx 70\%$) of
the volume in which antineutrinos are counted, the subsequent improvement in
energy resolution and gamma ray feature definition (Fig.~\ref{fig:th_spectrum})
allow for a considerably more reliable determination of its position. This is a
more important consideration for this detector, since we are more concerned
with the stability of our analysis energy thresholds than with having a
calibration that is averaged over the entire analyzed volume. Since we have no
indication that the relative optical transport between the different volumes
sampled by the calibration and the more general analysis has altered we are
confident that it is consistent to apply this calibration to all events.

Once the ADC bin locations of both the pedestal and $^{208}$Tl gamma ray
feature are determined, an absolute calibration is imposed. The nominal energy
of the $^{208}$Tl gamma ray differs from the measured value in the detector due
to energy leakage and other effects. To set the absolute energy scale, we
performed a Monte Carlo simulation of gamma transport in the detector, as
described in Sec.~\ref{sec:MC}.  The location of the $^{208}$Tl gamma ray
feature is expected to be at $2.39$~MeV. The error in this calibration
procedure results in an error of $\approx27~$keV  at $2.39~$MeV and
$\approx40~$keV at $3.5~$MeV, which are the energy thresholds used in later
analysis for prompt and delayed events, respectively. The calibrated PMT values
for a cell are added to determine the energy of events in that cell in later
analysis.

\section{Antineutrino Event Selection}
\label{sec:analysis}

As described above, the $10$ data items comprising an event are recorded
whenever both PMTs attached to a single cell produce a signal exceeding the
hardware thresholds. These are the ADC values for each of the 8~PMTs and both
the time since the last acquisition trigger and the time since either the last
acquisition trigger or the last muon veto hit, which ever is less. Events are
streamed to disk soon after they are recorded.

Events are written to disk sequentially, with a new file created every hour.
All analysis operations are performed on this unit of data. The product of this
analysis is a reduced data set that can be combined across many hours, as the
task at hand requires, and transmitted efficiently over a low bandwidth
datalink for remote monitoring purposes.


\subsection{Preliminary Analysis}

As described in Sec.~\ref{sec:nubarint}, we are looking for two energy
depositions, arising from the positron and neutron interactions, in close time
coincidence. The first step in the analysis is to form candidate event pairs
from the raw singles data. For every pair of sequential  events, the first is
labeled ``prompt,'' or positron-like, and the second is labeled ``delayed,'' or
neutron-like. Two time intervals are also recorded: the time before the prompt
event that the most recent muon veto or acquisition trigger occurred and the
time between the prompt and delayed events (interevent time).

A complete hour of data taking typically comprises $\approx~1.4$~million
prompt/delayed pairs. Several items of interest are derived from this complete
data set:
\begin{itemize}

\item{the total number of acquisition triggers recorded in the
hour,}

\item{the interevent time spectrum of the entire data set is histogramed, and fit to
an exponential to determine the acquisition trigger rate for that hour,}

\item{the calibration constants for the PMT channels are determined
as described in Sec.~\ref{sec:calibration}.}

\end{itemize}

The next step in the analysis is to reduce the data set by applying an energy
cut to the event pairs. To collect data for the automated calibration the
discriminator thresholds are set well below the energy of the $2.6$~MeV
$^{208}$Tl background line. However, as will be discussed below, the final
antineutrino analysis thresholds must be set higher, since the background rate
is too high at lower energies. Thus, a loose precut is made that requires
greater than $2.0$~MeV in both the prompt and delayed portions of an event,
reducing the size of the data set that must be subsequently analysed by more
than an order of magnitude.


\subsection{Event Selection}
\label{sec:event_selection}

To select antineutrino candidates we apply several further cuts.

\begin{figure}[tb]
\centering
\includegraphics*[width=3in]{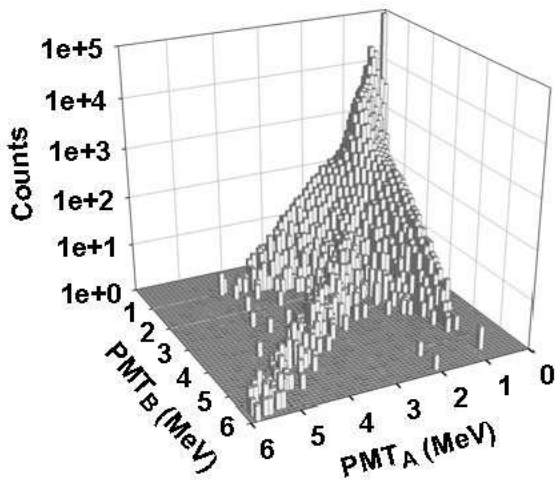}
\caption{Counts as a function of the share of energy in each PMT.}
\label{fig:ratio1}
\end{figure}

In order to remove a strong position dependency in the light collection
efficiency in our scintillator cells, a cut is applied to the ensure that each
PMT in a cell sees a similar amount of light per event. An example of the light
intensity seen by each PMT in a cell is shown in Fig.~\ref{fig:ratio1}. Two
clear features can be seen -- a line along $PMT_A=PMT_B$, where an event occurs
in the lower or middle portion of the cell so that the PMTs observe similar
amounts of light, and large numbers of events on the ``wings'' where the light
distribution is uneven. In those latter events the interaction point is closer
to one PMT than the other resulting in uneven light collection, as well as a
higher overall light collection efficiency that results in an incorrect energy
assignment.

\begin{figure}[tb]
\centering
\includegraphics*[width=3in]{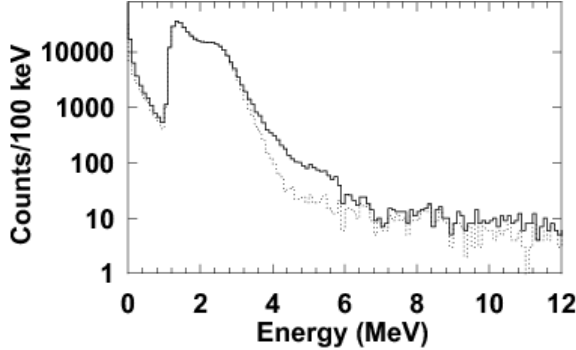}
\caption{The energy spectrum of a scintillator cell before (solid) and after
(dotted) the PMT ratio cut is applied.} \label{fig:ratio2}
\end{figure}

For this reason, we remove events with an uneven light distribution. This cut
is applied to the normalized ratio of the two calibrated PMT intensities,
\begin{equation}
r = \left|\frac{PMT_A-PMT_B}{PMT_A+PMT_B}\right| < 0.4,
\end{equation}
so that when the two PMTs observe similar intensity this quantity is small. The
effect of this cut on the energy spectrum of singles in a cell is demonstrated
in Fig.~\ref{fig:ratio2}. Many events with apparent high energies are removed
-- in reality these are low energy singles that occur close to one PMT,
yielding a higher light collection than events in the bulk of the scintillator
cell. Thus, the effect of this cut is to exclude a volume close to the PMTs.

A GEANT4~\cite{GEANT} simulation was used to estimate the size of the remaining
fiducial volume. Optical photons were generated throughout the cell volume, the
number reaching each PMT recorded, and the PMT ratio cut applied. The
simulation was repeating using several values for the scintillator attenuation
length and cell wall reflectivity consistent with the LED scan shown in
Fig.~\ref{fig:LEDfig}. The result was quite insensitive to these changes, and
it is estimated that events in $83\pm4\%$ of the cell volume will pass this
cut.

Next, we apply energy cuts to the amount of energy recorded in each cell. In
this analysis we consider each scintillator cell to be an independent detector,
i.e. we require that both the prompt and delayed energy distributions exceed an
analysis threshold, and do not consider simultaneous energy depositions in any
other cell. Because both the prompt and delayed energy depositions from an
inverse beta interaction include gamma rays, energy can in fact be transported
to adjacent scintillator cells, particularly for the delayed event (gamma ray
shower following neutron capture on Gd). However, the energy sharing effect is
relatively small -- we have determined that considering the energy sum of all
cells in the delayed energy determination would add only $\approx 10$\% to the
antineutrino count rate. This comes at the expense of increased complexity and
that all scintillator cell/PMT combinations must be fully operational during
the entire data taking period. For simplicity and flexibility in our analysis,
we have chosen to consider each cell as an independent detector.

\begin{figure}[t]
\centering
\includegraphics*[width=3in]{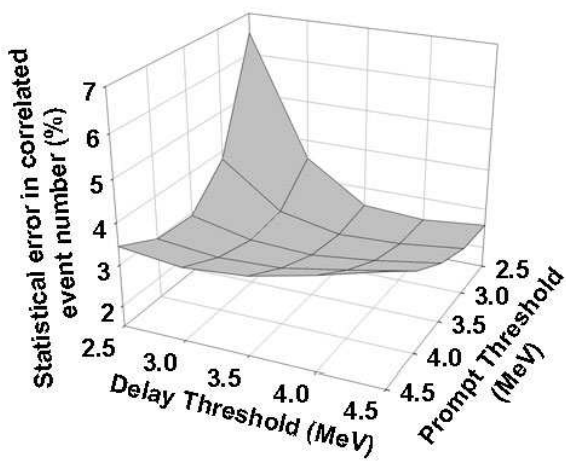}
\caption{The statistical uncertainty in the correlated event rate as a function
of the prompt and delayed thresholds.} \label{fig:threshold_optimization}
\end{figure}

The lower energy thresholds for prompt and delayed events are optimized with
respect to the long term stability of the threshold and the signal to
background ratio. A lower threshold admits more of the desired antineutrino
events but increases the number of uncorrelated background events, whose
subtraction introduces considerable statistical uncertainty. In the two
parameter space described by the prompt and delayed thresholds there is a
fairly broad minimum in the statistical uncertainty in the number of correlated
events. (Fig.~\ref{fig:threshold_optimization}). This gives us some freedom to
choose thresholds values based on the stability criterion.

For the prompt event the lower threshold is set at the clearly identifiable
$2.39$~MeV calibration feature. This reduces the dependence of the analysis on
the energy calibration, since no extrapolation of the energy scale from the
measured calibration point is required. The delayed energy threshold is set at
$3.5~$MeV. With the prompt threshold set as  just described, this value for the
delayed threshold is optimized with respect to minimizing statistical
uncertainty and maximizing the neutron detection efficiency.

Upper thresholds are chosen based on physical considerations. An upper
threshold of $9$~MeV is chosen for the prompt events since there are expected
to be very few energy depositions due to reactor antineutrinos with energy
greater than this (Sec.~\ref{sec:positron}). For the delayed event, an upper
threshold of $10$~MeV is chosen, based on the expected delayed energy spectrum
determined via Monte Carlo simulations (Sec.~\ref{sec:neutron_mc}). This value
reflects the maximum energy released by neutron capture on Gd, blurred by the
energy resolution of the scintillator cells.

\begin{figure}[tb]
\centering
\includegraphics*[width=3in]{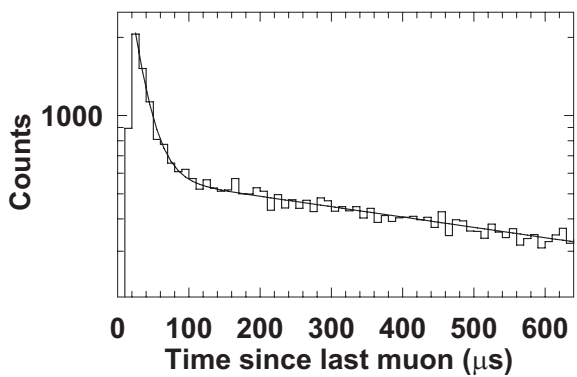}
\caption{Measured values of the time elapsed since the last muon or cell
event.} \label{fig:muon_spectrum}
\end{figure}

Finally, we apply a cut that suppresses antineutrino-like event pairs caused by
muon interactions. Fig.~\ref{fig:muon_spectrum} displays the spectrum of time
since the last muon or acquisition trigger values once all the cuts described
above have been applied. As expected for a Poisson process, the dominant
feature is a simple exponential with time constant equal to inverse the rate at
which events occur, in this case the sum of the muon veto hit rate and the
acquisition trigger rate. At shorter times there are additional events that are
correlated with muon hits in the veto -- it is these that we must be careful to
eliminate. For this reason we apply a cut on this value, considering only those
events that occur at least $100~\mu$s after the last muon hit/acquisition
trigger. From the measured time constant of the muon correlated events
($20~\mu$s), we calculate that this cut excludes all but $0.7\%$ of events
correlated with a muon recorded in the veto.

\subsection{Antineutrino count rate determination}
\label{sec:nubar_count_rate_determination}

\begin{figure}[tb]
\centering
\includegraphics*[width=3in]{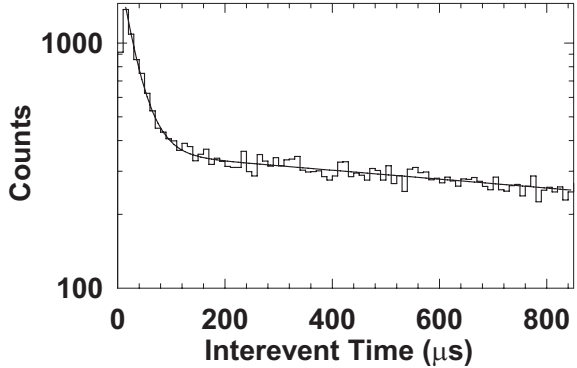}
\caption{A representative interevent time spectrum of event pairs that pass all
selection cuts acquired during $7~$days of data taking.}
\label{fig:t12_spectrum}
\end{figure}

After applying these cuts, we then examine the interevent time spectrum
(Fig.~\ref{fig:t12_spectrum}). Two clear exponential features are visible. The
first and most interesting is that due to correlated pairs of events, a
positron-like energy deposition followed a characteristic time later by the
capture of a neutron on a Gd nucleus. The second exponential is that due to
random coincidences of sequential background events that both exceed the
relevant thresholds. As would be expected for such event pairs, the time
constant of this exponential is equal to the acquisition trigger rate.

Because it is impossible in this detector to distinguish random coincidences
from true correlated event pairs on an event-by-event basis, we use a
statistical separation of the two to determine the correlated
(antineutrino-like) event rate. This is achieved by fitting the spectrum
displayed in Fig.~\ref{fig:t12_spectrum} to the sum of two exponentials. The
two time constants are known in advance; the uncorrelated rate is measured from
the raw data, and the neutron capture time constant is measured from a high
statistics data set ($28.6\pm0.3~\mu$s). Thus, we need only fit for two
parameters, the amplitude of each exponential. Simple integrals,
\begin{equation}
N=A\int_{10~{\mu}s}^{800~{\mu}s}e^{-t/\tau}\,dt ,
\end{equation}
then yield the number of event pairs that fall into each class. The fit is
implemented using the CERN MINUIT package~\cite{MINUT}, using full error
propagation, so that bin-to-bin statistical errors are correctly applied in
determining the best fit exponential amplitudes and their uncertainties.

\begin{figure}[tb]
\centering
\includegraphics*[width=3in]{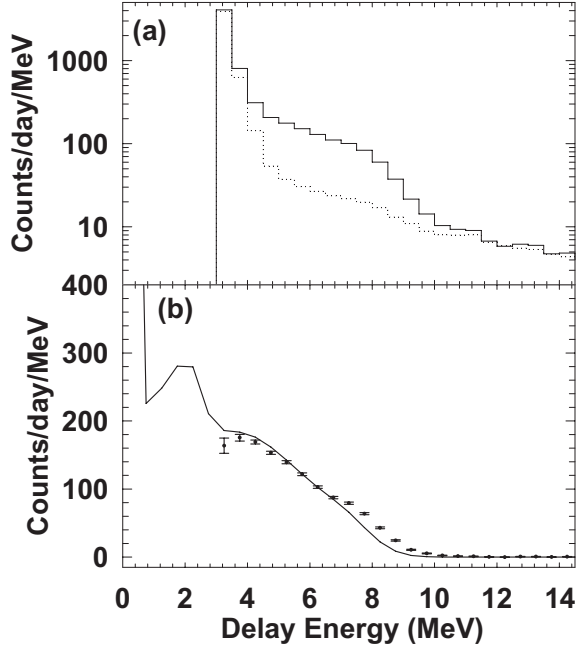}
\caption{Energy spectra of correlated events (delayed in this example) are
obtained using a statistical separation. (a) The average energy spectrum of
uncorrelated events is determined using events with long interevent time
(dotted) and is then subtracted from spectra determined at short interevent
times (solid). This yields the average energy spectrum of the correlated events
(b). The solid line shows the neutron energy spectrum predicted by the Monte
Carlo, including the expected feature at $\approx 2.2~$MeV due to capture on
hydrogen.} \label{fig:e_spectrum_extraction}
\end{figure}

The energy spectra of the prompt and delayed portions of correlated events can
also be extracted using a statistical separation -- this is again necessary due
to our inability to distinguish correlated and uncorrelated events on an
event-by-event basis. The procedure used is demonstrated for the delayed case
in Fig.~\ref{fig:e_spectrum_extraction}. The average energy spectra for
uncorrelated events are determined from events with long interevent times
($400-800~\mu$s), while the sum of the average energy spectra for correlated
and uncorrelated events is determined from those with short interevent times
($10-100~\mu$s). The uncorrelated spectra are scaled so that there are equal
numbers of uncorrelated events in both the short and long time energy spectra,
and are then subtracted from the short time spectra, yielding the average
correlated energy spectra.

\section{Confirmation of Reactor Antineutrino Detection}

Having extracted the number of correlated events, we then wish to determine
what fraction of those are due to antineutrino interactions. Unfortunately we
have no means to distinguish these from correlated backgrounds, such as those
caused by fast neutrons. Instead, we must wait for the antineutrino source, the
SONGS Unit~$2$ reactor, to turn off. This occurs approximately every
$1.5~$years during regularly scheduled refueling outages, as well as
occasionally for inspections or brief unscheduled maintenance operations. With
the Unit~$2$ reactor off the detector counts correlated background events and a
small number of antineutrino interactions from the relatively distant Unit~$3$.

\begin{figure}[tb]
\centering
\includegraphics*[width=3in]{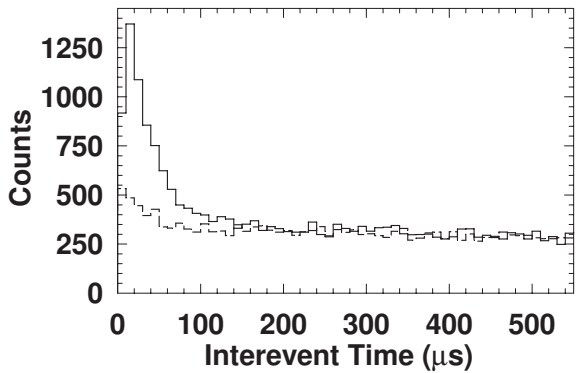}
\caption{Interevent time spectra acquired over $7$ days with the Unit~$2$
reactor off (dashed) and on (solid). } \label{fig:t12_on_off}
\end{figure}

Fig.~\ref{fig:t12_on_off} displays the interevent time spectra for two $7$~day
periods in April and May of 2006. In the first period, Unit~$2$ was at zero
power for refueling. During this time around $100$ correlated events pass all
selection cuts. In the second, refueling was completed and the reactor was
operating at full power.

Clearly, there are many more correlated events when the reactor is on, compared
to when the reactor is off. Also, the uncorrelated portion of the interevent
time spectrum is unaltered by the change in reactor state, as are the detector
trigger and muon veto rate. The number of correlated and uncorrelated events
contained in the two spectra is compared in Table~\ref{tab:on_off_results}. The
reactor off measurement is essentially a measurement of the correlated
background rate. It comprises less than $20\%$ of the correlated event rate,
giving good signal to noise for a measurement of the reactor antineutrino rate.
The $7$~day average of the observed antineutrino detection rate,
$459\pm16$~/day, is in good agreement with that predicted, $407\pm75$~/day.

\begin{table}
\caption{Comparison of measurements with the Unit~2 reactor on and off.}
\label{tab:on_off_results}
\begin{tabular}{l r r} \hline
Reactor State       & Off               & On \\ \hline
Uncorrelated Events &~$3732\pm25$~/day~        &  ~$3785\pm26$~/day      \\
Correlated Events   &~$105\pm9$~/day        &  ~$564\pm13$~/day      \\
DAQ trigger rate    &~$420\pm1$~Hz           &  ~$418\pm1$~Hz         \\
Muon veto rate      &~$506\pm1$~Hz          &  ~$503\pm1$~Hz         \\ \hline
\end{tabular}
\end{table}

\begin{figure}[tb]
\centering
\includegraphics*[width=3in]{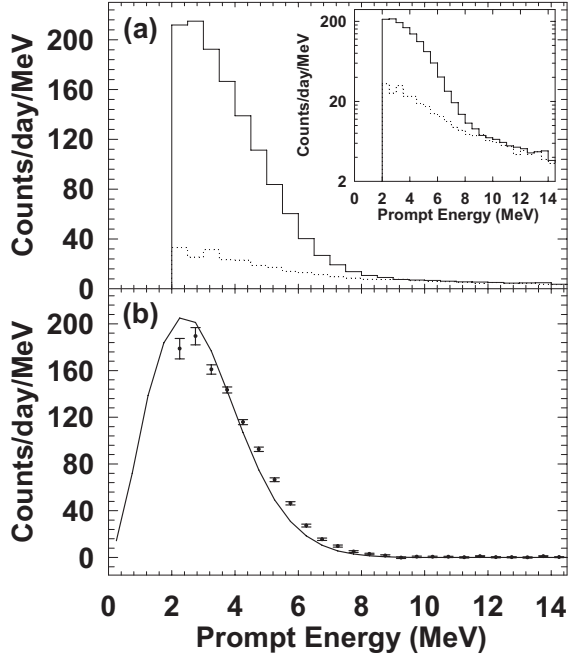}
\caption{(a) The energy spectrum of prompt  correlated events with the Unit~2
reactor on (solid) and off (dotted). The logarithmic insert clearly shows the
agreement between the two spectra at energies greater than $9$~MeV. (b) The
inferred inverse beta positron energy spectrum derived by subtracting the
reactor off data from the reactor on data. The solid line shows the Monte Carlo
prediction, including transport effects and the simple optical model.}
\label{fig:prompt_energy}
\end{figure}

To further confirm that we are observing reactor antineutrinos, we can also
examine the prompt event energy spectrum during reactor on and off periods
(Fig.~\ref{fig:prompt_energy}a). With the Unit~2 reactor off, we are measuring
the prompt energy spectrum of the correlated backgrounds, while with the
reactor on we measure the sum of these backgrounds and the prompt antineutrino
interaction energy deposition.

As discussed in Sec.~\ref{sec:nubarint} the prompt energy deposition in an
inverse beta interaction is essentially that of the incoming antineutrino. The
expected prompt energy spectrum was given in Sec.~\ref{sec:positron}; we expect
the distribution to go to zero at around $10$~MeV. This is what we observe in
Fig.~\ref{fig:prompt_energy}a -- above this energy the reactor on and reactor
off distributions agree.

Furthermore, we can determine the prompt energy spectrum of the reactor
correlated events (Fig.~\ref{fig:prompt_energy}b), and compare it to that
expected from the Monte Carlo calculation (Sec.~\ref{sec:positron}). The
agreement is good, again confirming that we are indeed detecting reactor
antineutrinos.

In principle, the highest energy neutrons in the reactor spectrum could create
a correlated antineutrino-like signal in the detector in that same way that
fast neutrons produced by cosmic rays do. However, as noted in
Sec.~\ref{sec:site}, this rate is negligible, since reactor fast and thermal
neutrons from the core are completely attenuated by the $\approx20~$m of
concrete between the core and the detector. The reactor associated gamma flux
at the detector location is similarly negligible.

\section{Conclusion}
\label{sec:conclusion}

Our deployment of an antineutrino detector at a nuclear reactor has shown that
even a cubic meter scale detector of modest efficiency can easily measure
hundreds of antineutrino events per day, a large enough count rate to be of
interest for cooperative monitoring regimes. Of particular interest is the
relative smallness of the detector compared to many other modern reactor
antineutrino detectors. (These earlier detectors have typically used larger
active volumes to improve containment of the spatially diffuse neutron-induced
gamma shower, while we have sacrificed some efficiency to reduce our overall
footprint.) We have also shown that the detector can operate for months to
years with only very occasional maintenance needed by the safeguards inspector.
This is an important point since inspector-days dominate the cost of reactor
safeguards regimes. Calibrations can be performed using omnipresent gamma ray
backgrounds, without the need for external check sources. Concerning stability
of response, we have been able to use scintillator mixed several years prior to
our experiment, for two years running, with no observable reduction in response
due to aging. This perhaps surprising result again relates to the smallness  of
our detector, where the typical travel paths for scintillation photons are only
of the order $1$-$2$~meters compared to several meters or tens of meters in
other experiments. Our impact on day-to-day operations at the reactor site has
been negligible, an important consideration for gaining acceptance by reactor
operators. As far as the deployment site itself is concerned, as noted above in
Sec.~\ref{sec:site}, we are gratified to have found that many reactors around
the world appear to possess annular galleries with overburden and standoff
characteristics similar to the ones enjoyed in our own installation.

In summary, our initial experience with deployment augurs well for the
near-term utility of this novel reactor safeguards tool. Shortly, we will also
publish analyses of our current data set quantifying the precision with which
thermal power and fissile content can be measured with the SONGS1 design.

\begin{ack}
We thank DOE NA-22 for their continued support of this project. We are indebted
to the management and staff of the San Onofre Nuclear Generating Station for
site access and the excellent support they have provided us. N.~Madden,
D.~Carr, J.~Gollnick, and A.~Salmi made essential contributions to the
electronic and mechanical design and construction of the detector. We thank
D.~McNabb for assistance with ``dicebox'' calculations. We are particularly
grateful to F.~Boehm for providing us with the liquid scintillator used in this
detector. We gratefully acknowledge useful discussions with R.~Svoboda and
S.~Dazeley.
\end{ack}

\end{document}